\documentclass[11pt,
 amsmath,amssymb,
 aps,
prb,
]{revtex4-2}

\usepackage{graphicx}
\usepackage{bm}
\usepackage{amsmath}
\usepackage{tabularx}
\usepackage{float}

\usepackage{hyperref}
\hypersetup{
    colorlinks=true,
    linkcolor=red,
    urlcolor=blue,
    citecolor=blue
}

\usepackage{array}
\newcolumntype{L}[1]{>{\raggedright\let\newline\\\arraybackslash\hspace{0pt}}m{#1}}

\begin{document}

\makeatletter
\def\frontmatter@abstractheading{} 
\makeatother


\title{Dislocation Dynamics and Shape in High Entropy Alloys: the Influence of Stress Correlations, Long-Range Interactions and Anisotropy}

\author{D\'enes Berta}
\author{P\'eter Dus\'an Isp\'anovity}
\affiliation{ELTE E\"otv\"os Lor\'and University, Department of Materials Physics, P\'azm\'any P\'eter s\'et\'any 1/A, 1117 Budapest, Hungary}

\begin{abstract}
High entropy alloys gained significant scientific interest in recent years due to their enhanced mechanical properties including high yield strength combined with outstanding ductility. The strength of these materials originates from their highly heterogeneous pinning stress fields that hinder dislocation glide, that is, plastic deformation. This work investigates how the correlations and the anisotropy of the pinning stresses, and the long-range nature and the anisotropy of dislocation interactions influence the propagation of dislocations and the depinning transition in these alloys. Furthermore, it is studied how the impact of these factors manifest in the shape of dislocations. The implications to the wider scope of generic disordered systems and to possible experimental applications are also discussed.
\end{abstract}

\maketitle



\section{Introduction}

Elastic interfaces pinned by disordered media are ubiquitously present in nature. Examples include dislocation and crack propagation \cite{moretti2004depinning, ponson2006two, bonamy2008crackling, song2025enabling} domain wall dynamics  \cite{lemerle1998domain, zapperi1998dynamics, sethna2001crackling}, Barkhausen noise in ferromagnetic materials \cite{urbach1995interface, spasojevic1996barkhausen, mehta2002universal}, depinning of contact lines in wetting \cite{joanny1984model, moulinet2002roughness} and the stick-slip motion during tectonic activity and earthquakes \cite{gutenberg1956energy, fisher1997statistics}. Under external loading these systems exhibit intermittent motion and avalanche-like behavior, and the interface shape can become fractal-like. Statistical properties of their dynamics are often very similar (e.g., in terms of event size/strength distribution or spatial clustering) even if there are different mechanisms and time- and length-scales involved \cite{fisher1998collective, ispanovity2022dislocation, kuki2023statistical}.

One specific example (which is the narrower focus of this study) is the case of dislocation propagation in high entropy alloys (HEAs). These alloys consist of several (by convention, at least five) principal components, each contributing with the same (or similar) atomic concentrations. In these materials the combination of random atomic order and differences in atomic radii result in a quite heterogeneous pinning stress field. The stress fluctuations introduce effective obstacles (pinning centers) to dislocation motion \cite{utt2022origin, zhang2022data}, hence, they hinder plastic deformation. Thus, these materials have remarkably high yield strength which is combined with relatively high ductility. This combination, accompanied by other beneficial properties such as enhanced thermal stability and corrosion resistance \cite{tsai2014high}, raised great scientific interest about this class of materials and their possible applications. Therefore, the reliable modeling of depinning in HEAs can not only prove valuable in understanding depinning in disordered media generally, but can be important in enabling the optimization of the design of such materials.

In order to study plasticity on the level of individual dislocations in high entropy alloys (HEAs), medium entropy alloys (MEAs) or in random alloys generally, several frameworks have been employed, including atomistic simulations \cite{patinet2008depinning, patinet2011atomic, peterffy2020length, esfandiarpour2022edge} and modeling using a discrete dislocation approach both for dilute \cite{foreman1966dislocation, labusch1976movement, nogaret2006finite}  and concentrated alloys \cite{zapperi2001depinning, rida2022influence, song2025enabling}. Modeling dislocation behavior in disordered media has two key ingredients that should be captured appropriately: (i) the self-interaction of dislocations and (ii) the heterogeneity of the pinning stress field of the disordered medium. Dislocation self-interaction has two key features: long-ranged nature (that is, an interaction decaying according to a power-law) and anisotropy -- motivating the distinction of dislocations of different characters. While the long-range interaction of dislocation segments have been treated directly in several studies \cite{zapperi2001depinning, song2025enabling}, in others often line-tension approximations are applied \cite{geslin2018thermal, zhai2019properties, rida2022influence}. Regarding the pinning stress conditions, the two key features are the distribution of stresses and spatial correlations which have been shown to influence, e.g., the depinning transition \cite{park2003dynamics, patinet2013quantitative}.

This paper focuses on quantifying the impact of these key factors (long-range interaction, interaction anisotropy, stress fluctuations, and spatial stress correlations) on the depinning transition, dislocation roughening, and the statistics of individual dislocation slip events which has not been a central theme in most of the previous studies on dislocation behavior in alloys expect for a few examples \cite{ma2020unusual,   song2025enabling}. It is stressed that while this paper focuses on dislocation behavior in concentrated alloys, it has also implications for the more generic scope of understanding the behavior of elastic interfaces in disordered media.

\section{Numerical model}

For this work a two-dimensional dislocation model (previously applied in other studies \cite{zapperi2001depinning, song2025enabling}) is employed to simulate the motion of dislocations in the heterogeneous pinning stress field of a high entropy alloy. The model is constructed as follows. The simulation plane $(x,y)$ coincides with the slip plane of the dislocation under investigation. The inhomogeneous pinning stress field is defined on a lattice of size $L_x\times L_y$ where the linear cell size is chosen to be equal to the length $b$ of the Burgers vector. Periodic boundary condition (PBC) is applied in the direction $x$. The position of the dislocation is tracked as follows. Each cell is assigned a binary value: 1 if the cell has been already passed by the dislocation and 0 if not. Thus, the boundary between cells of different values represents the dislocation line, which consists of vertical and horizontal segments. In our simulations, the Burgers vector of the dislocation is chosen to be parallel either to the $x$ or $y$ axis, hence each dislocation segment has a purely edge or screw character. The motion of a segment (with a `center of mass') at position $\bm r$ is determined by the shear stress $\sigma(\bm r)$ acting on it, which is the sum of three contributions:
\begin{equation}
    \sigma(\bm r) = \sigma_\mathrm{ext}(\bm r)+\sigma_\mathrm{self}(\bm r)+\sigma_\mathrm{pin}(\bm r).
\end{equation}
Here, $\sigma_\mathrm{ext}$ is the externally applied stress, $\sigma_\mathrm{self}$ is the stress due to mutual interactions with other dislocation segments, and $\sigma_\mathrm{pin}$ represents the heterogeneous pinning stress field arising from the random atomic distribution inside the HEA. The simulations use a random dynamics approach: at each time step, a single segment is selected randomly (each having the same probability of being selected) and the segment is moved in the direction corresponding to the net shear stress acting on it. While this approach introduces randomness to the method, in previous works, it has been shown to produce physically correct results \cite{zapperi2001depinning, song2025enabling}. Below it is described in detail how the three stress components are determined in this model.

\subsection{Interaction of Dislocation Segments}

The self-stress $\sigma_\mathrm{self}$ acting on the $i$-th dislocation segment is given by:
\begin{equation}
    \sigma_\mathrm{self}(\bm r_i)=\sum_j^{\{s\}}\sigma_\mathrm{s}(\bm r_i-\bm r_j)+\sum_j^{\{e\}}\sigma_\mathrm{e}(\bm r_i-\bm r_j)
\end{equation}
Here, $\sigma_\mathrm{s}$ and $\sigma_\mathrm{e}$ are the shear stress fields of screw and edge segments, respectively, and $\{s\}$ and $\{e\}$ are the sets of screw and edge segments from which the $i$th segment is excluded. The PBC in the direction $x$ is taken into account using periodically placed `images' of the `real' dislocation segments inside the simulation cell. These images are also included in the $\{s\}$ and $\{e\}$ sets. Since the stress field of the segments decays relatively quickly (as $\frac{1}{r^2}$), only 3 images of each segment are included on each side. The shear stress due to screw and edge segments is given by
\begin{align}
    \sigma_\mathrm{s}(\bm r)&=S_\mathrm{s}\frac{\mu b^2 r_\parallel}{4\pi r^3},
    \label{eq:stress_screw}
    \\
    \sigma_\mathrm{e}(\bm r)&=S_\mathrm{e}\frac{\mu b^2 r_\perp}{4\pi(1-\nu) r^3}.
    \label{eq:stress_edge}
\end{align}
Here, $r$ is the magnitude of the relative position vector $\bm r$ measured from the segment. $r_\parallel$ and $r_\perp$ are the components of $\bm r$ parallel and perpendicular to the Burgers vector, respectively. $\nu$ is the Poisson ratio, $\mu$ is the shear modulus, and $b$ is the magnitude of the Burgers vector $\bm b$. $S_\mathrm{s}=\pm 1$ and $S_\mathrm{e}=\pm1$ are dimensionless prefactors, with the sign depending on the line direction of the segments.

\subsection{Pinning Stress Field}

The stress field $\sigma_\mathrm{pin}$ representing the inhomogeneous pinning stress of the HEA is generated as random noise. Based on micro-elastic models and the central limit theorem, we can expect the pinning stress field distribution to be approximately Gaussian \cite{geslin2021microelasticityI}. Consequently, normally distributed $\sigma_\mathrm{pin}$ is generated, with its standard deviation denoted by $\Sigma_\mathrm{pin}$. The simplest assumption is that this noise is uncorrelated (at the resolution $b$ used in the model). For part of the simulations, this assumption is followed and spatially uncorrelated Gaussian $\sigma_\mathrm{pin}$ fields are generated. However, the previously mentioned micro-elastic models also indicate that the pinning shear stress field shows anisotropic spatial correlation \cite{geslin2021microelasticityII}, which influences the propagation of dislocations \cite{rida2022influence}. Let the autocorrelation of the $\sigma_{ij}$ component of the pinning stress field be denoted by $\tilde{\mathcal{A}}_{ij}$, where the tilde highlights that the quantity is taken in Fourier space. The autocorrelation in Fourier space is given by \cite{geslin2021microelasticityII}:
\begin{equation}
    \tilde{\mathcal{A}}_{ij}(\bm k)=\frac{120\pi^\frac{3}{2}b^3}{L_xL_yL_z}\frac{k_i^2k_j^2}{k^4}e^{-b^2k^2}.
\end{equation}
Here, $b$ is the magnitude of the Burgers vector, $\bm k$ is the wave vector in Fourier space, and $L_x$, $L_y$, and $L_z$ are the dimensions of the medium in the $x$, $y$, and $z$ directions, respectively. For the given autocorrelation, the correlated $\sigma_\mathrm{pin}$ stress field is obtained by the following procedure \cite{rida2022influence}. The random stress field is first determined in Fourier space as:
\begin{equation}
    \tilde{\sigma}_\mathrm{pin}(\bm k)=\sqrt{\frac{\tilde{\mathcal{A}}_{ij}(\bm k)}{2}}\left(\mathcal N_\mathrm{re}+\mathrm{i}\mathcal N_\mathrm{im}\right)
\end{equation}

Here, $\mathrm{i}$ is the imaginary unit, and $\mathcal N_\mathrm{re}$ and $\mathcal N_\mathrm{im}$ are normally distributed random numbers. Finally, the inverse Fourier transform of the result yields the stress field $\sigma_\mathrm{pin}(\bm r)$ in real space.

\subsection{External Loading}

In numerical methods of dislocation dynamics simulations typically either pure stress control \cite{ispanovity2014avalanches, lehtinen2016glassy, berta2023dynamic, berta2024avalanche, berta2025identifying} or pure strain control \cite{agnihotri2015rate, fan2021strain, kurunczi2021dislocation, kurunczi2023avalanches} is implemented. However, due to the finite stiffness of the specimens and the devices these protocols cannot be realized in experimental setups. In the model used in this paper, therefore, an intermediate approach is utilized that falls between these two limiting cases. At each time step of the simulations, the external stress $\sigma_\mathrm{ext}$ (which is considered to be uniform across the entire lattice) is updated as follows. On the one hand, the stress is increased at a fixed rate: in every time step, a constant value $\Delta \sigma$ is added to the current stress. In addition, the plastic deformation resulting from the motion of the dislocation causes a drop in stress due to the elastic response of the sample and the device. Since in our model only a single dislocation segment moves during each time step, the swept area during a time step (which is proportional to the increment of the plastic deformation) is $\Delta A = \pm b^2$, where the sign depends on the direction of the motion. The change in the external stress due to the plastic deformation is given by $-B\frac{\Delta A}{b^2}=\mp B$, where $B$ characterizes the stiffness of the system. Denoting the external stress after the $i$-th time step by $\sigma_\mathrm{ext}^{(i)}$, and the signed area swept during that step by $\Delta A^{(i)}$, we have
\begin{equation}
    \sigma_\mathrm{ext}^{(i)}=\sigma_\mathrm{ext}^{(i-1)}+\Delta\sigma-B\frac{\Delta A^{(i)}}{b^2}.
\end{equation}

\subsection{Parameter Selection}

The model parameters were chosen as follows. As it was mentioned above, the size of the simulation domain is $L_x \times L_y$, measured in units of $b$. The domain height $L_y$ is fixed at $400b$ in our simulations. Since the length of the Burgers vector is typically $1.5$–$2~\textup{\AA}$, this corresponds to approximately $80~\mathrm{nm}$. The domain width was varied as $L_x = 50b, 100b, 150b, 200b$ (i.e., roughly within the range of $10$–$40~\mathrm{nm}$) in order to observe and filter out possible size effects of the model.

According to classical elasticity theory, the Poisson ratio can vary in the range $-1 < \nu < 0.5$, but in real materials its typical range is $0.2 < \nu < 0.5$ \cite{mott2009limits}. In our simulations, we chose the Poisson ratio as $\nu = 0.35$ which is similar to the typical values for MEAs and HEAs \cite{laplanche2015temperature, laplanche2020processing, jha2023phase}.

This work aims to study the depinning transition (at the critical stress $\tau_\mathrm{crit}$) and the shape and dynamics of dislocations near the critical state. Therefore, the stress is changed slowly, and the stiffness of the medium is tuned in such manner that the resulting stress drops are of a magnitude that allows the system to remain near the critical state even after the dislocation motion has initiated. These conditions are satisfied if
\begin{equation}
    \Delta\sigma \ll B \ll \tau_\mathrm{crit}.
\end{equation}
The chosen parameter values under these conditions are $\Delta\sigma = 10^{-10}\mu$ and $B = 10^{-8}\mu$, where $\mu$ is the shear modulus, which also defines the unit of stress in our simulations since the interaction stresses scale with $\mu$ [see Eqs.~(\ref{eq:stress_screw}) and (\ref{eq:stress_edge})]. For HEAs, $\mu \simeq 100~\mathrm{GPa}$, hence these parameters translate to $\Delta \sigma \simeq 10~\mathrm{Pa}$ and $B \simeq 1~\mathrm{kPa}$.

\begin{figure}[H]
    \centering
    \includegraphics[width=0.9\linewidth]{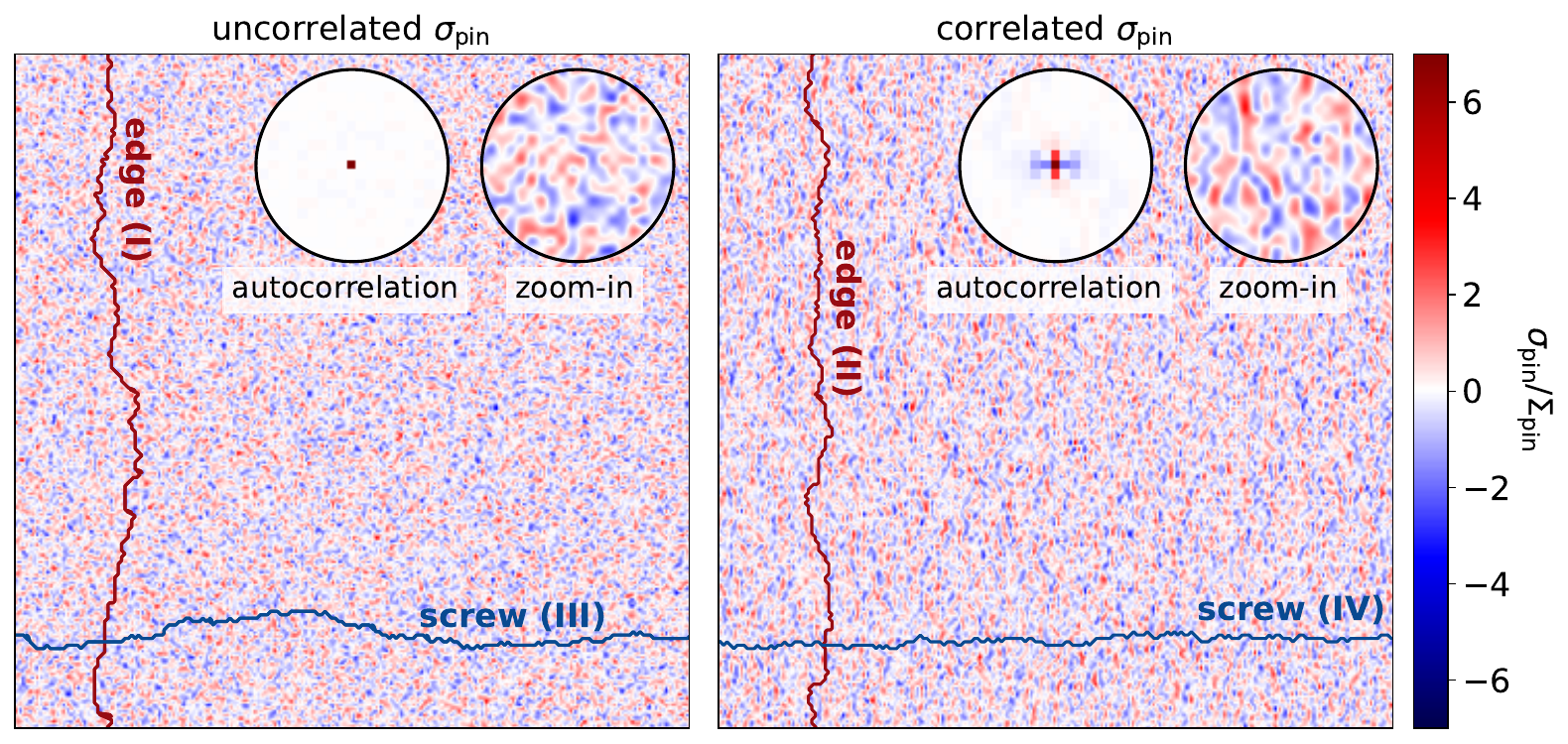}
    \caption{\textbf{The four studied cases.} Cases I and III correspond to edge and screw dislocations, respectively, moving in uncorrelated pinning stress field $\sigma_\mathrm{pin}$. Cases II and IV correspond to edge and screw dislocations, respectively, propagating in correlated stress field. The circular insets show the autocorrelation of the pinning stress field and zoom into the pinning stress field. $\Sigma_\mathrm{pin}$ denote the standard deviation of $\sigma_\mathrm{pin}$. Note that $\sigma_\mathrm{pin}$ has been interpolated for this visualization below the resolution $b$ that is used in our simulations.}
    \label{fig:cases}
\end{figure}

The standard deviation of the pinning stress was tuned as $\Sigma_\mathrm{pin} = 0.0175\mu$ in a previous work to mimic a specific HEA (the so-called Cantor alloy) \cite{song2025enabling}. Thus, the values used in the present study are from the range $0.005\mu\leq\Sigma_\mathrm{pin}\leq 0.05\mu$, which corresponds to approximately $0.5~\mathrm{GPa} < \Sigma_\mathrm{pin} < 5~\mathrm{GPa}$. For our statistical analysis, we generated configurations for each $(L_x, \Sigma_\mathrm{pin})$ combination, with 5 configurations for 4 distinct cases. These cases (illustrated in Fig.~\ref{fig:cases}) are the following:
\begin{itemize}
    \item[(I)] Edge dislocation moving in uncorrelated pinning stress field $\sigma_\mathrm{pin}$,
    \item[(II)] Edge dislocation moving in correlated pinning stress field $\sigma_\mathrm{pin}$,
    \item[(III)] Screw dislocation moving in uncorrelated pinning stress field $\sigma_\mathrm{pin}$,
    \item[(IV)] Screw dislocation moving in correlated pinning stress field $\sigma_\mathrm{pin}$.
\end{itemize}
We stress that since the dislocations are not perfectly straight in the heterogeneous medium, of course, the are not purely of edge or screw character in either case -- only at the very beginning of the simulations.

\section{Results}

\subsection{Depinning Transition}

\begin{figure}[H]
    \centering
    \includegraphics[width=0.5\linewidth]{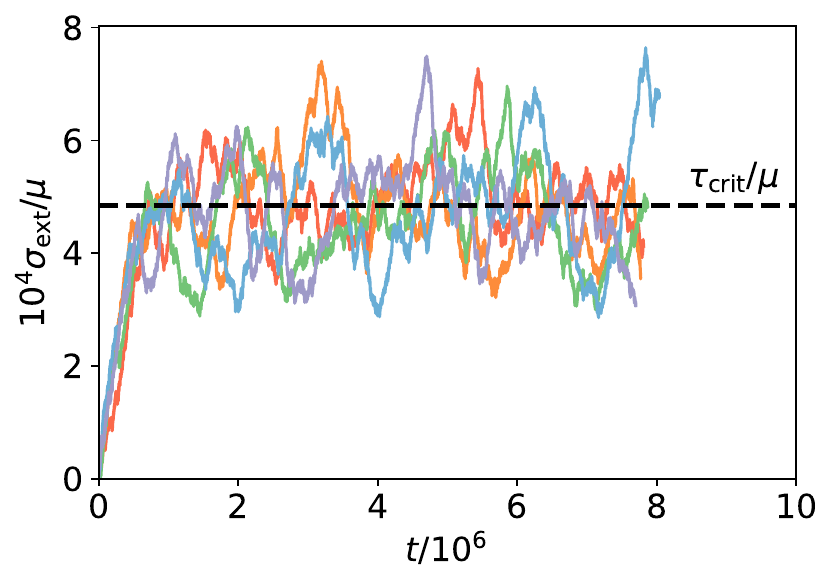}
    \caption{\textbf{Critical stress for different realizations.} Stress-time curves of edge dislocations moving in an uncorrelated pinning stress field $\sigma_\mathrm{pin}$ with a standard deviation of $\Sigma_\mathrm{pin}=0.0051\mu$, in systems of width $L_x=200b$. $t$ denotes the number of elapsed time steps, $\sigma_\mathrm{ext}$ is the external stress, $\mu$ is the shear modulus, and $\tau_\mathrm{crit}$ is the critical stress corresponding to the depinning transition. The curves in different colors represent five different realizations. The dashed line shows the average critical stress. Its uncertainty is estimated using the Jackknife method: $\tau_\text{crit}=(4.84\pm0.05)\cdot10^{-4}\mu$. Note that despite the relatively large stress fluctuations, the mean critical stress can be estimated with high precision.}
    \label{fig:tau_crit_example}
\end{figure}

To determine the critical stress $\tau_\mathrm{crit}$ corresponding to the depinning transition, an initially straight (pure edge or pure screw) dislocation is placed in the simulation domain along the direction $x$. Starting from a zero initial external stress $\sigma_\mathrm{ext}$, the stress is slowly increased based on the previously described loading procedure. After the onset of intermittent dislocation motion, the stress-time curve becomes jerky. The critical stress $\tau_\mathrm{crit}$ is identified as the average stress over multiple realizations in this intermittent regime. This intermittent behavior is shown in Fig.~\ref{fig:tau_crit_example} where the dashed line indicates the critical stress which can be determined with high precision relative to the magnitude of stress fluctuations.

\begin{figure}[H]
    \centering
    \includegraphics[width=0.9\linewidth]{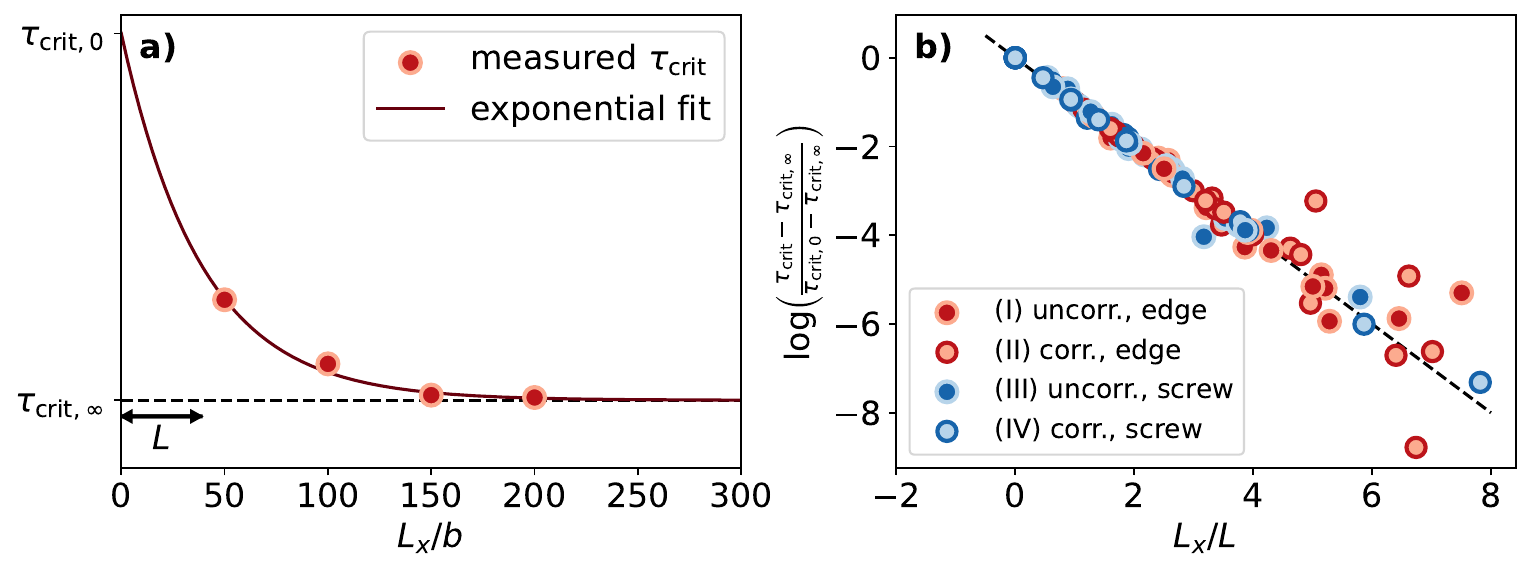}
    \caption{\textbf{Non-physical size-effect in the critical stress.} (a): The critical stress $\tau_\mathrm{crit}$ for edge dislocations moving in an uncorrelated pinning stress field $\sigma_\mathrm{pin}$ with standard deviation $\Sigma_\mathrm{pin}=0.0051\mu$ ($\mu$ being the shear modulus) across different system widths $L_x$. $b$ denotes the length of the Burgers vector. The critical stress approaches the asymptotic value $\tau_{\mathrm{crit},\infty}$ exponentially, which corresponds to the limit $L_x\to\infty$. The other limiting case ($L_x\to 0$) is characterized by $\tau_{\mathrm{crit},0}$. The exponential convergence is characterized by the length scale $L$, which is also indicated in the figure. (b): All investigated systems (differing in terms of correlation, dislocation character, and $\Sigma_\mathrm{pin}$) exhibit exponential convergence of $\tau_\mathrm{crit}$. In the case of exponential convergence, the data points lie along the dashed line.}
    \label{fig:tau_crit_size_effect}
\end{figure}

\begin{figure}[H]
    \centering
    \includegraphics[width=0.45\linewidth]{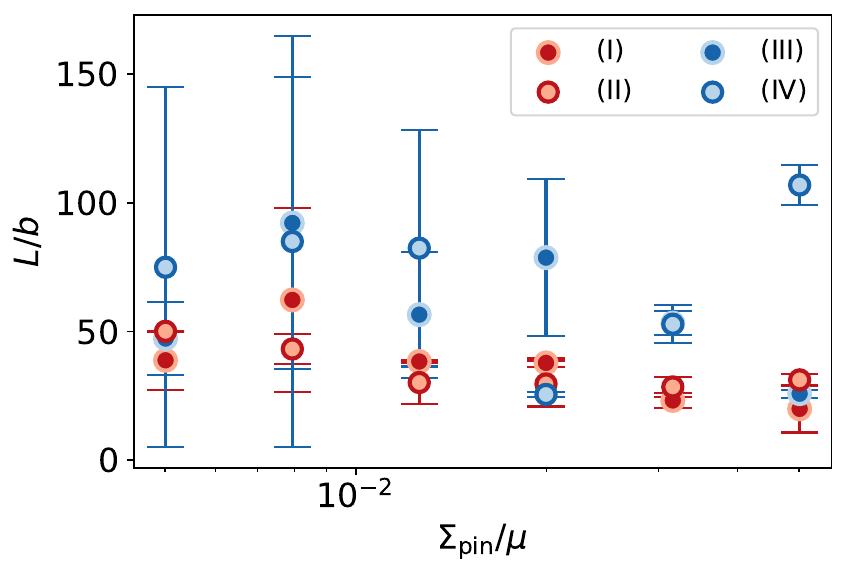}
    \caption{\textbf{Characteristic length of the convergence of the critical stress.} The characteristic length $L$, defined in Eq.~(\ref{eq:tau_crit_convergence}), for different dislocation characters and under pinning stress fields with varying standard deviations $\Sigma_\mathrm{pin}$ and spatial correlations. $b$ denotes the length of the Burgers vector, and $\mu$ is the shear modulus.}
    \label{fig:tau_crit_convergence}
\end{figure}

Upon examining the value of $\tau_\mathrm{crit}$ while varying dislocation character, $\Sigma_\mathrm{pin}$, and the presence or lack of spatial stress correlations across various system widths $L_x$, the following is observed. As demonstrated in Fig.~\ref{fig:tau_crit_size_effect}(a) for the specific case of an edge dislocation moving in an uncorrelated $\sigma_\mathrm{pin}$ field with standard deviation $\Sigma_\mathrm{pin}=0.0051\mu$, as the system width increases the critical stress exponentially approaches an asymptotic value. Thus, formally
\begin{equation}
\tau_\mathrm{crit}=\tau_{\mathrm{crit},\infty}+(\tau_{\mathrm{crit},0}-\tau_{\mathrm{crit},\infty})e^{-L_x/L},
\label{eq:tau_crit_convergence}
\end{equation}
where $\tau_{\mathrm{crit},0}$ and $\tau_{\mathrm{crit},\infty}$ denote the critical stresses in the limits $L_x\to0$ and $L_x\to\infty$, respectively, while $L$ represents the characteristic length of the convergence towards the asymptotic value $\tau_{\mathrm{crit},\infty}$. By rearranging Eq.~(\ref{eq:tau_crit_convergence}), we obtain:
\begin{equation}
\mathrm{log}\left(\frac{\tau_\mathrm{crit}-\tau_{\mathrm{crit},\infty}}{\tau_{\mathrm{crit},0}-\tau_{\mathrm{crit},\infty}}\right)=-\frac{L_x}{L}.
\end{equation}
Figure~\ref{fig:tau_crit_size_effect}(b) demonstrates that this relationship (indicating exponential convergence with the system width) holds not only in the specific case shown in Fig.~\ref{fig:tau_crit_size_effect}(a) but for all examined system types. Similar exponential convergence has been observed in other dislocation models with periodic boundary conditions \cite{rida2022influence, vaid2022pinning}, however, it is important to stress that this is not a physical size effect (due to finite specimen size), but rather a result of the finite simulation domain. This convergence property of the critical stress enables us to extrapolate and to analyse the asymptotic $\tau_{\mathrm{crit},\infty}$ value representative of infinitely large systems that do not carry artificial finite-size effects. We also note that for the largest systems we examined ($L_x=200b$), the values of $\tau_\mathrm{crit}$ and $\tau_{\mathrm{crit},\infty}$ are extremely close, since the characteristic length $L$ is generally much smaller than $200b$, as shown in Fig.~\ref{fig:tau_crit_convergence}. The figure also indicates that the properties of the $\sigma_\mathrm{pin}$ stress field do not have a very clear effect on the value of $L$. However, it is observed that the typical value of $L$ differs depending on dislocation character: for edge dislocations $L/b=37\pm11$, and for screw dislocations $L/b=65\pm25$. According to Eq.~\ref{eq:tau_crit_convergence}, this results in a typical deviation from the asymptotic critical stress of only $0.4\%$ for edge dislocations and $5\%$ for screw dislocations for the largest systems examined.

\begin{figure}[H]
    \centering
    \includegraphics[width=0.5\linewidth]{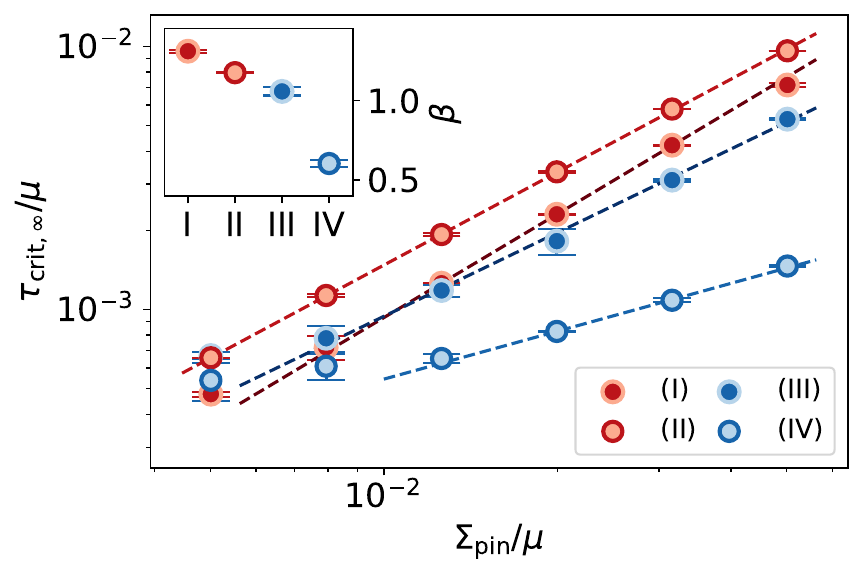}
    \caption{\textbf{Dependence of the critical stress on the pinning stress fluctuations.} The asymptotic critical stress $\tau_{\mathrm{crit},\infty}$ of infinitely wide systems scales with the standard deviation $\Sigma_\mathrm{pin}$ of the pinning stress field. The inset shows the scaling exponent $\beta$ for the four system types. $\mu$ denotes the shear modulus.}
    \label{fig:tau_crit_scaling}
\end{figure}

Figure \ref{fig:tau_crit_scaling} shows that for a given dislocation character and spatial correlation type, the critical stress scales with the standard deviation $\Sigma_\mathrm{pin}$ of the pinning stress field, i.e.,
\begin{equation}
\tau_{\mathrm{crit},\infty}\propto \Sigma_\mathrm{pin}^\beta,
\label{eq:beta}
\end{equation}
where $\beta$ is the scaling exponent. It is evident that the dislocation character, the the fluctuations and the correlations of the pinning stresses all influence the critical stress in the following ways: (i) for edge dislocations, the exponent $\beta$ is higher than for screw dislocations, (ii) when correlations in the $\sigma_\mathrm{pin}$ field are considered, the value of $\beta$ decreases, which is particularly pronounced for screw dislocations, (iii) within the examined range (relevant for HEAs), the highest critical stress values correspond to edge dislocations in correlated fields, while the lowest values are taken by screw dislocations in correlated fields, and (iv) the influence of dislocation character is significantly greater in correlated $\sigma_\mathrm{pin}$ fields. Observation (iv) can be understood simply by noting that the spatial correlations are different for edge and for screw dislocations, which amplifies the importance of the dislocation character during dislocation motion.

\subsection{Dislocation Shape}

\begin{figure}[H]
    \centering
    \includegraphics[width=\linewidth]{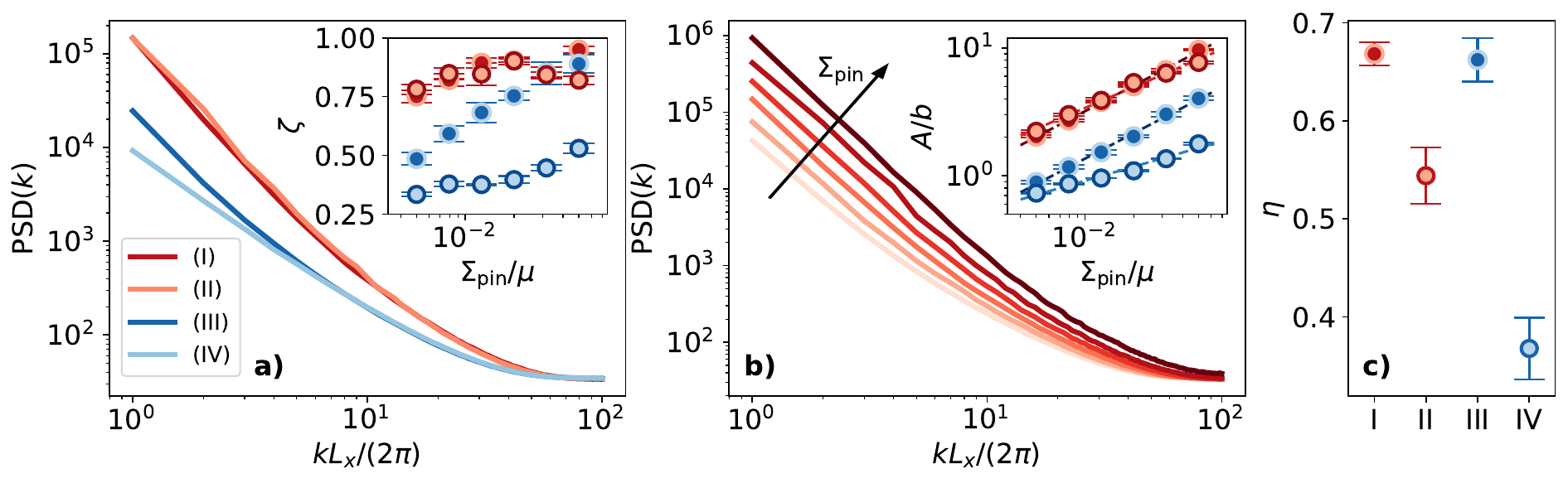}
    \caption{\textbf{Power spectral density of dislocation lines.} (a) Average power spectral density $\mathrm{PSD}(k)$ of dislocations in the critical state in systems of width $L_x=200b$, where the standard deviation of the pinning stress is $\Sigma_\mathrm{pin}=0.01259\mu$. $\mu$ is the shear modulus, and $k$ is the wavenumber of the Fourier modes. The inset shows the evolution of the exponent $\zeta$, defined in Eq.~(\ref{eq:zeta}), as a function of $\Sigma_\mathrm{pin}$. (b) Dependence of the power spectral density on $\Sigma_\mathrm{pin}$ for edge dislocations moving in uncorrelated fields (i.e., case I) in systems of width $L_x=200b$. The inset shows the $\Sigma_\mathrm{pin}$ dependence of the amplitude $A$, defined by Eq.~(\ref{eq:A}). $b$ denotes the length of the Burgers vector. (c) The exponent $\eta$, defined in Eq.~(\ref{eq:eta}), in various types of systems.}
    \label{fig:PSD}
\end{figure}

Previous numerical studies have shown that dislocations have a fractal-like shape in the inhomogeneous pinning fields of alloys \cite{zapperi2001depinning,zhai2019properties,peterffy2020length}. This shape can be characterized by the roughness exponent $\zeta$. For a dislocation propagating in the direction $y$,
\begin{equation}
    \langle |y(x+d)-y(x)|\rangle_x\propto d^\zeta,
    \label{eq:roughness_def}
\end{equation}
where $\langle\bullet\rangle_x$ denotes averaging over $x$ and $d$ is a distance in the direction $x$. Instead of using the definition in (\ref{eq:roughness_def}), the exponent $\zeta$ is often determined from the Fourier spectrum of the dislocation line. Let $k$ be the wavenumber of the Fourier modes and $\mathrm{PSD}(k)$ their power spectral density. In the case of a fractal-like shape,
\begin{equation}
    \mathrm{PSD}(k)\propto k^{-\zeta'},
    \label{eq:spectral}
\end{equation}
from which the exponent $\zeta$ can be determined as \cite{priol2021long, song2025enabling}
\begin{equation}
    \zeta=\frac{\zeta'-1}{2}.
    \label{eq:zeta}
\end{equation}
To characterize the magnitude of the spatial fluctuations of the dislocation line, we can introduce the quantity
\begin{equation}
    A=\frac{2b}{L_x}\sqrt{\mathrm{PSD}(2\pi/L_x)}
    \label{eq:A}
\end{equation}
which quantifies the typical amplitude of the longest wavelength fluctuations. Based on Fig.~\ref{fig:PSD}, the relation in Eq.~(\ref{eq:spectral}) indeed holds for dislocation lines, although the spectral densities differ quantitatively across different system types. The quantities $\zeta$ and $A$ are also influenced by the standard deviation of the pinning stress field. The $\zeta$ exponent typically increases with increasing $\Sigma_\mathrm{pin}$ and appears to saturate around $\zeta = 0.9$ based on the available data [see inset of Fig.~\ref{fig:PSD}(a)]. The extent of the dislocation line fluctuations (and thus $A$) also depends on the value of $\Sigma_\mathrm{pin}$ [see Fig.~\ref{fig:PSD}(b)]
\begin{equation}
    A\propto\Sigma_\mathrm{pin}^\eta
    \label{eq:eta}
\end{equation}
The value of $\eta$ in different systems is shown in Fig.~\ref{fig:PSD}(c).

\subsubsection*{Dislocation Dynamics}

In the following, the stick-slip dynamics in the critical state is analysed. Due to the slow loading, the dislocation slip events are well separated in time, allowing for easy identification and distinction.

\begin{figure}[H]
\centering
\includegraphics[width=0.9\linewidth]{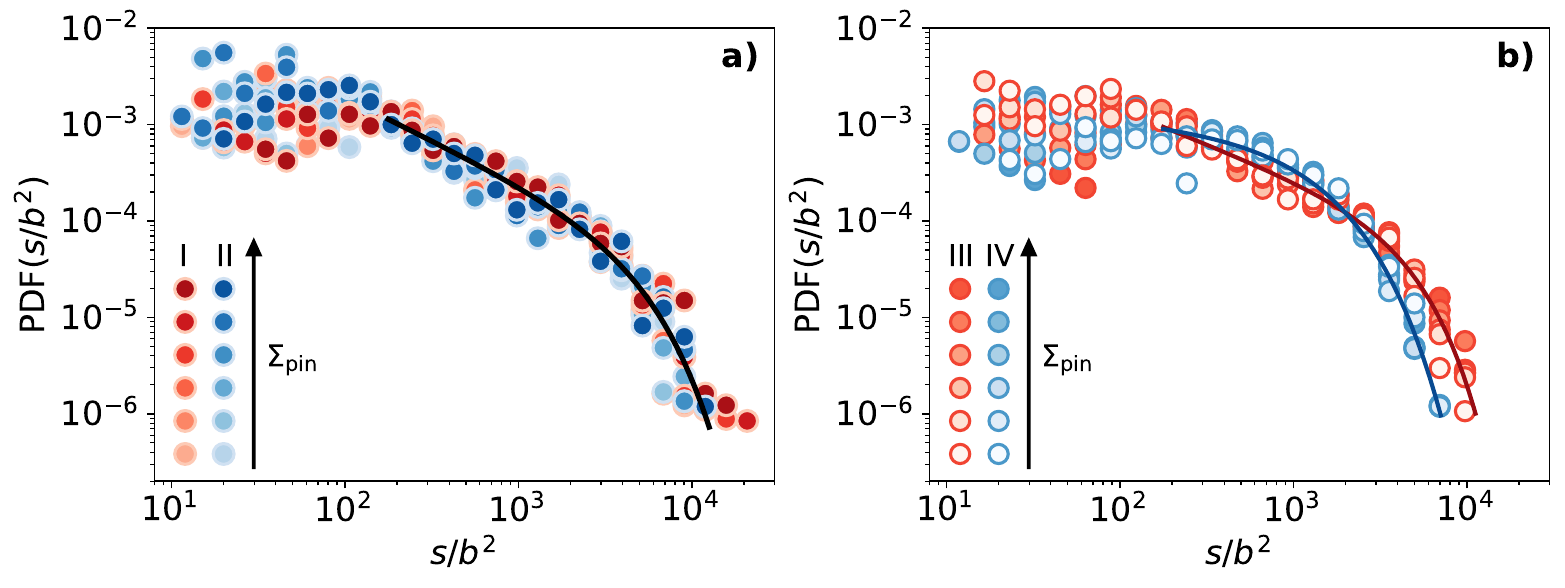}
\caption{\textbf{Size distribution of dislocation slip events.} The probability density function (PDF) of the swept area $s$ of dislocation slip events. $\Sigma_\mathrm{pin}$ denotes the standard deviation of the pinning stress field. The system width is $L_x=200b$, where $b$ is the magnitude of the Burgers vector. (a) and (b) show the slip distribution in uncorrelated and correlated pinning stress fields, respectively.}
\label{fig:event_size}
\end{figure}

Let $s$ denote the area swept by a single event. Based on previous results, we expect that the probability density function PDF of $s$ has the form of a power-law with a streched exponential cut-off, that is,
\begin{equation}
\mathrm{PDF}(s)\propto s^{-\tau}e^{-(s/s_0)^\gamma},
\label{eq:s_PDF}
\end{equation}
where $\tau$, $\gamma$ and $s_0$ are constants, the last characterising the (soft) upper limit of the event size \cite{song2025enabling}. The size distribution of the events was examined in the largest systems with a width of $L_x=200b$. As shown in Fig.~\ref{fig:event_size}(a), the function in Eq.~(\ref{eq:s_PDF}) fits well for dislocations moving in uncorrelated fields. Regardless of the dislocation character and the value of $\Sigma_\mathrm{pin}$, $\tau=0.9\pm0.1$, $s_0=(4600\pm900)b^2$, and $\gamma=1.3\pm0.2$ are found. The data suggest that the cut-off length $s_0$ slightly depends on the dislocation character and the magnitude of stress fluctuations $\Sigma_\mathrm{pin}$. The latter dependence can be formulated as
\begin{equation}
s_0\propto\Sigma_\mathrm{pin}^\delta,
\end{equation}
where $\delta=0.30\pm0.05$.

In contrast to the uncorrelated case, for a correlated stress field $\sigma_\mathrm{pin}$, the data do not indicate any $\Sigma_\mathrm{pin}$ dependence [see Fig.~\ref{fig:event_size}(b)]. However, the distribution is clearly influenced by the dislocation character. For edge dislocations, we find $\tau=0.7\pm0.1$, $s_0=(4400\pm1000)b^2$, and $\gamma=1.4\pm0.3$. For screw dislocations moving in a correlated field, the cut-off length $s_0=(1100\pm300)b^2$ is so small that for a given system size, there is no clear scale-free region, and practically only the cut-off regime is fitted, which makes the other fitting parameters unreliable. Accurate measurement of $\tau$ and $\gamma$ would require significantly wider systems (which demand more computational resources).

\begin{figure}[H]
\centering
\includegraphics[width=0.9\linewidth]{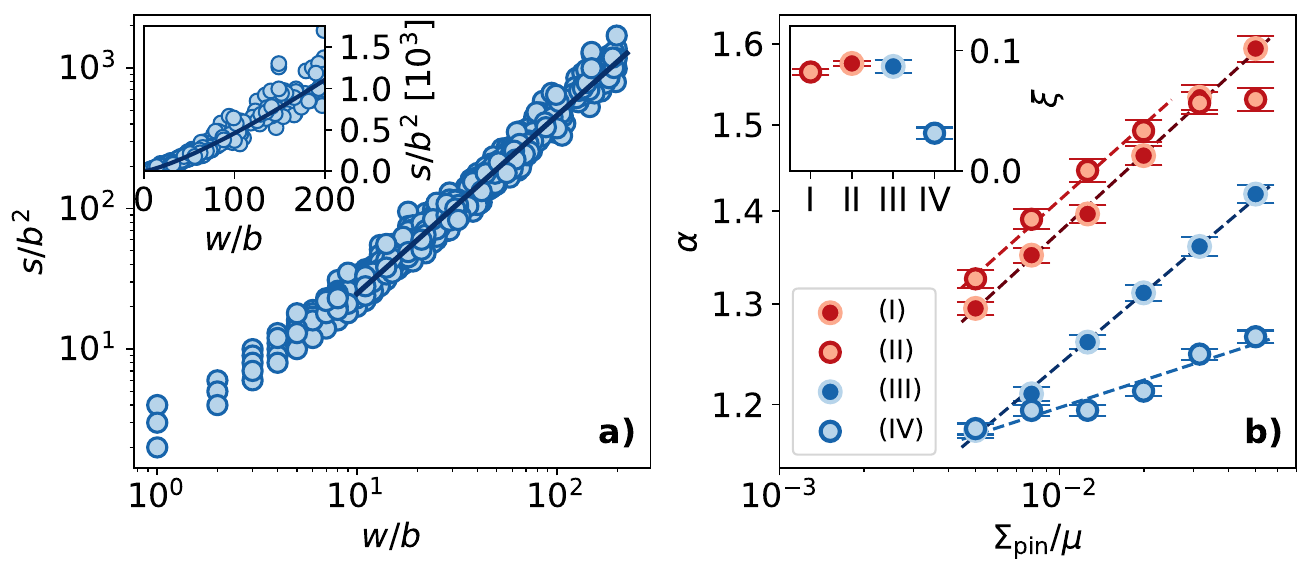}
\caption{\textbf{The shape of dislocation slip events.} (a) The scaling between the width $w$ and area $s$ of slip events for screw dislocations moving in a correlated stress field (i.e., case IV) with standard deviation $\Sigma_\mathrm{pin}=0.05012\mu$. $b$ denotes the magnitude of the Burgers vector and $\mu$ is the shear modulus. Here, data from all system sizes were aggregated. The inset shows the scaling on linear scales. (b) The $\alpha$ exponent defined in equation Eq.~(\ref{eq:shape_scaling}) and shown as the slope in panel (a) as a function of the standard deviation $\Sigma_\mathrm{pin}$ of the pinning stress field. The exponent $\xi$ characterizing the scaling [see Eq.~(\ref{eq:xi})] is shown in the inset.}
\label{fig:event_shape}
\end{figure}

Let $w$ denote the width (extension in direction $x$) of the events. Our previous results indicated that the relation between the width and area of the events follows the scaling
\begin{equation}
s\propto w^\alpha
\label{eq:shape_scaling}
\end{equation}
characterized by a nontrivial fractal dimension $1<\alpha<2$ \cite{song2025enabling}. Our current analysis shows that the value of $\alpha$ depends on the dislocation character and the characteristics of the stress field $\sigma_\mathrm{pin}$. The scaling described by Eq.~(\ref{eq:shape_scaling}) is observed in all types of systems considered in this study, and is shown for a specific type of system in Fig.~\ref{fig:event_shape}(a). In Fig.~\ref{fig:event_shape}(b), it is also shown that the exponent $\alpha$ weakly but clearly depends on the standard deviation $\Sigma_\mathrm{pin}$ of the pinning stress field. This dependence obeys
\begin{equation}
\alpha\propto \Sigma_\mathrm{pin}^\xi,
\label{eq:xi}
\end{equation}
where $\xi$ is a constant characteristic to the tpye of system considered. Note that the system width $L_x$ did not affect the value of the exponent $\alpha$, therefore, for each tpye of system, data from all system sizes were used to determine the exponent. For screw dislocations moving in a correlated field, $\xi\approx0.03$, while in all other cases, $\xi\approx0.085$. It can also be clearly stated that $\alpha$ is larger for edge dislocations. This difference can be interpreted based on a simple energetical argumentation. The field of a screw segment is smaller by a factor of approximately $(1-\nu)\approx2/3$ than that of an edge segment ($\nu$ being the Poisson number that is typically around $1/3$) , meaning that the bending out (involving the creation of edge segments) of a screw dislocation during a slip event is energetically less favourable than in the case of an edge dislocation, where the bending out results in the formation of screw segments of lower energy contribution.

\section{Discussion}

In this section, our results are compared to previous results from the literature obtained by employing various dislocation models and more generic models of interface dynamics in disordered media. These include lattice-based dislocation dynamics models the same as or similar to ours (LDD), other discrete dislocation dynamics studies (DDD), molecular dynamics studies (MD), the quenched Edqard-Wilkonson model (qEW), the quenched Kardar-Parisi Zhang model (qKPZ), Ginzburg-Landau type models (GL), random bond models (RB), elastic models (EM), line tension approximations (LTA) and classical force-based models (CFM). Tables \ref{tab:comparison1}, \ref{tab:comparison2} and \ref{tab:comparison3} summarise the relevant parameters available for comparison. Note that in cases where a more generic model do not correspond to either edge or screw dislocation case the parameter values are listed in both rows. Furthermore, some studies also included spatial correlation in the pinning stress fields, however, not necessarily with the same autocorrelation. The possible differences are not emphasised in the tables.

Generally it is observed that the parameters obtained in this work fit well in the picture drawn by the parameter ranges of other similar models. However, one striking difference is that while the present results suggest that in correlated stress fields the exponent $\beta$ is lower for screw dislocations ($\beta\approx0.6$), it is shown in Ref.~\cite{rida2022influence} to be significantly higher (well above $1$) for screw dislocations compared to their edge counterparts. The possible reason for this difference is that while our model considers long-range interactions between dislocation segments, the other models in which values for $\beta_\mathrm{screw}^\mathrm{corr}$ are available apply a line tension type, short-range approximation. Another inconsistency can be noted, in the case of the roughness exponent $\zeta^\mathrm{corr}_\mathrm{screw}$. For that parameter our value is significantly lower than the one in Ref.~\cite{zapperi2001depinning}. It difference may originate from the fact that while that study employs the same dislocation model (in fact, it motivated our study), the pinning stresses are correlated in an isotropic manner, different from our case.

\begin{table}[H]
\centering
\begin{tabular}{cccccccc}
\hline\hline
     & This work & EM \cite{rida2022influence} & LTA \cite{rida2022influence}  & CFM \cite{rida2022influence} & DDD \cite{zhai2019properties} & qEW \cite{ferrero2013nonsteady}
    \\ \hline\hline
    $\beta_\mathrm{edge}^\mathrm{uncorr}$ & $1.31\pm0.01$ & -- & $\approx1.33$ & $4/3$ & $1.00-1.33$ & $4/3$ 
    \\ 
    $\beta_\mathrm{screw}^\mathrm{uncorr}$ & $1.06\pm0.03$ & -- & $\approx1.33$ & $4/3$ &  --  & $4/3$ 
    \\
    $\beta_\mathrm{edge}^\mathrm{corr}$ & $1.18\pm0.01$ & $1.20\pm0.01$ & $1.20\pm0.01$ & $4/3$ &  -- &  --
    \\ 
    $\beta_\mathrm{screw}^\mathrm{corr}$ & $0.60\pm0.02$ & $2.42\pm0.02$ & $1.33-2.73$ & $2$ &  -- &  --
 \\ \hline\hline
\end{tabular}
\caption{The exponent $\beta$ characterizing the scaling of the critical stress $\tau_\mathrm{
crit}$ of the depinning transition with the standard deviation $\Sigma_\mathrm{pin}$ for different models [for the definition see Eq.~(\ref{eq:beta})]. The upper and lower indices indicate the presence or lack or pinning stress correlations and dislocation character, respectively.}
\label{tab:comparison1}
\end{table}

\begin{table}[H]
\centering
\begin{tabular}{ccccccccc}
\hline\hline
     & This work &  DDD \cite{zhai2019properties} & LDD \cite{zapperi2001depinning} & MD \cite{peterffy2020length} & qEW \cite{kim2006depinning} & qKPZ \cite{mukerjee2023depinning} & GL \cite{kolton2023depinning}
    \\ \hline\hline
    $\zeta_\mathrm{edge}^\mathrm{uncorr}$ & $0.75-0.95$ & $\approx1$ & --  & $0.69-0.93$ & $1.250\pm0.003$ & $\approx0.63$ & $0.5-1.2$
    \\
    $\zeta_\mathrm{screw}^\mathrm{corr}$ & $0.33-0.53$ &  -- & $\approx1$ & -- & -- & -- & --
 \\ \hline\hline
\end{tabular}
\caption{The exponent $\zeta$ characterizing the roughness of dislocation lines at the critical state for different models [for the definition see Eq.~(\ref{eq:zeta})]. The upper and lower indices indicate the presence or lack or pinning stress correlations and dislocation character, respectively.}
\label{tab:comparison2}
\end{table}

\begin{table}[H]
\centering
\begin{tabular}{cccccc}
\hline\hline
     & This work &  LDD \cite{song2025enabling} & RB \cite{rosso2009avalanche} & qEW \cite{aragon2016avalanches} & qKPZ \cite{mukerjee2023depinning}
    \\ \hline\hline
    $\tau^\mathrm{uncorr}$ & $0.9\pm0.1$ & $\approx1$ & $1.08\pm0.02$ & $\approx1.11$ & $\approx1.26$
    \\
    $\alpha^\mathrm{uncorr}$ & $1.2-1.6$ & $1.3-1.4$ & -- & -- & --
 \\ \hline\hline
\end{tabular}
\caption{The exponents $\tau$ and $\alpha$ characterizing the size distribution and shape of dislocation slip events for different models with uncorrelated pinning stress fields [for the definitions see Eqs.~(\ref{eq:s_PDF}) and (\ref{eq:shape_scaling})].}
\label{tab:comparison3}
\end{table}

\section{Conclusion}

In this work, the influence of the pinning field correlation, long-range interactions and anisotropy in pinning and dislocation stress fields was studied. In particular, the depinning transition, dislocation shape and the stick-slip dislocation dynamics was investigated which were all affected by these factors. The critical stress (characterizing the depinning transition) was shown to depend strongly on the pinning stress field of the HEA. The dislocation character alone does not influence criticality significantly, however, the combined effect of the dislocation character and the pinning stress anisotropy has immense impact on the critical stress. The dislocation shapes (at the critical state) were also shown to be affected by the studied factors, however, the shape of edge dislocations do not seem to depend much on pinning stress correlation. In all four studied cases the power spectral density of dislocation lines obeys a power law with the parameters of the spectrum dependent on dislocation character and the pinning stress field. It was demonstrated that the size of individual dislocation slip events obeys a power law distribution with a stretched exponential cut-off. In an uncorrelated pinning stress field, the distribution seems to be independent of the dislocation character, whereas if pinning stress correlations are introduced, the dynamics of edge and screw dislocations are visibly different. The slip events were demonstrated to exhibit a fractal-like behavior in terms of their shape with a fractal dimension between 1 and 2. For screw dislocations, the fractal dimension was shown to be lower (closer to 1) than for edge dislocations, indicating that the events are flatter on average for screw dislocations.

All these results indicate that it is important to take into account the pinning stress correlations when studying dislocation behavior in alloys. Generally, dislocation character did not affect much the behavior in uncorrelated stress field implying that the anisotropy of the self-interaction is less relevant. On the one hand, our results generally fit well in the zoo of dislocation and interface dynamics models. On the other hand, differences to other models (see Tab.~\ref{tab:comparison1}) and the observation that the relevant exponents are typically not universal but vary as the parameters and properties of the model are changed. These findings accentuate that the details of such models have to be chosen carefully because they might affect the results (e.g., the value of the critical stress). It is noted that while the present work focused on the behavior of dislocations in HEAs, however, the results can be indicative to the more generic scope of interface dynamics in disordered media.

\section*{Acknowledgments}
D.B. was supported by the EKÖP-24 University Excellence Scholarship Program of the Ministry for Culture and Innovation from the source of the National Research, Development and Innovation Fund. Financial support from the National Research, Development and Innovation Fund of Hungary under the young researchers’ excellence program NKFIH-FK-138975 is also acknowledged (P.D.I.).


\bibliographystyle{naturemag}
\bibliography{HEA}

\end{document}